\documentclass[aps,prb,reprint,groupedaddress]{revtex4-2}
\usepackage{graphicx}% Include figure files
\usepackage{bm}% bold math
\usepackage{color}
\usepackage{amsmath}

\begin{document}

%\preprint{}

\title{Crystal field excitation in the chiral helimagnet YbNi$_3$Al$_9$}


\author{Mitsuru Tsukagoshi}
%\thanks{}
\affiliation{Department of Quantum Matter, ADSE, Hiroshima University, Higashi-Hiroshima 739-8530, Japan}
\author{Suguru Kishida}
\affiliation{Department of Quantum Matter, ADSE, Hiroshima University, Higashi-Hiroshima 739-8530, Japan}
\author{Kenshin Kurauchi}
\affiliation{Department of Quantum Matter, ADSE, Hiroshima University, Higashi-Hiroshima 739-8530, Japan}
\author{Daichi Ito}
\affiliation{Department of Quantum Matter, ADSE, Hiroshima University, Higashi-Hiroshima 739-8530, Japan}
\author{Koya Kubo}
\affiliation{Department of Quantum Matter, ADSE, Hiroshima University, Higashi-Hiroshima 739-8530, Japan}
\author{Takeshi Matsumura}
\email[]{tmatsu@hiroshima-u.ac.jp}
\affiliation{Department of Quantum Matter, ADSE, Hiroshima University, Higashi-Hiroshima 739-8530, Japan}
\author{Yoichi Ikeda}
\affiliation{Institute for Materials Research, Tohoku University, Sendai, 980-8577, Japan}
\author{Shota Nakamura}
\affiliation{Department of Physical Science and Engineering, Graduate School of Engineering, Nagoya Institute of Technology, Nagoya 466-8555, Japan}
\author{Shigeo Ohara}
\affiliation{Department of Physical Science and Engineering, Graduate School of Engineering, Nagoya Institute of Technology, Nagoya 466-8555, Japan}


\date{\today}

\begin{abstract}
Crystal field level scheme of a uniaxial chiral helimagnet YbNi$_3$Al$_9$, exhibiting a chiral magnetic soliton lattice state by Cu substitution for Ni, has been determined by inelastic neutron scattering. 
The ground and the first excited doublets are separated by 44 K and are simply expressed as $\alpha|\pm 7/2\rangle + \beta |\mp 5/2\rangle$ with $\alpha$ and $\beta$ nearly equal to $\pm 1/\sqrt{2}$. 
The easy axis of the crystal field anisotropy is the $c$ axis when the excited levels are populated at high temperatures and high magnetic fields. 
On the other hand, the magnetism at low temperatures and low magnetic fields, where only the ground doublet is populated, is described by an easy plane anisotropy which may be treated as an $S=1/2$ system with an anisotropic $g$-factor, $g_{xy}=3.02$ and $g_z=1.14$. 
An orbital dependent exchange interaction is also discussed to explain the temperature dependence of the magnetic susceptibility based on this level scheme. 
\end{abstract}

%\keywords{}

\maketitle

\section{Introduction}
\label{sec:Introduction}
In chiral magnets with neither a space inversion center nor a mirror reflection plane, Dzyaloshinskii-Moriya antisymmetric interaction arises in addition to the normal symmetric exchange interaction, resulting in a helimagnetic structure with a unique sense of rotation. 
This effect gives rise to a nontrivial magnetic structure in magnetic fields applied perpendicular to the helical propagation vector $\bm{q}$ in uniaxial helimagnets, where $\bm{q}$ is parallel to the principal axis in the tetragonal, hexagonal, or trigonal crystal systems. 
This is called a chiral magnetic soliton lattice, a periodic array of helimagnetic kinks in a ferromagnetic background~\cite{Togawa12,Kishine15,Togawa16}. 
YbNi$_3$Al$_9$, belonging to a rhombohedral space group $R32$, has been attracting interest as a rare-earth helimagnet exhibiting such nontrivial magnetic structures. However, the $4f$ electron state of the Yb$^{3+}$ ion ($L$=3, $S$=1/2, $J$=7/2, $g$=8/7) in the crystalline electric field (CEF) of YbNi$_3$Al$_9$ has not been understood well. 

YbNi$_3$Al$_9$ exhibits a helical magnetic order below $T_{\text{N}}=3.4$ K with $\bm{q}=(0, 0, 0.82)$ with its magnetic moments lying in the $ab$-plane and propagate along the $c$-axis~\cite{Yamashita12,Miyazaki12,Tobash11,Utsumi12,Aoki18,Ota20,Wang20}. 
When a magnetic field is applied perpendicular to the $c$-axis, the helical order jumps to a forced ferromagnetic state at 0.1 T. 
In Cu substituted systems Yb(Ni$_{1-x}$Cu$_{x}$)$_3$Al$_9$, the critical field increases to 1.0 T and a chiral soliton lattice state is realized before the transition to the ferromagnetic state~\cite{Ohara14,Matsumura17,Ninomiya18}. 
For $H \parallel c$, the magnetization curve $M(H)$ exhibits a monotonic increase without a detectable anomaly at the transition from the expected conical state to the induced ferromagnetic state. 
These properties may suggest an easy plane anisotropy of the CEF ground state. 
It is remarked that, however, although $H \parallel c$ is a hard axis at low temperatures and low fields, $M_c(H)$ overtakes $M_{ab}(H)$ above $\sim 4$ T~\cite{Yamashita12,Ninomiya18,Ito20}. 
To proceed with further study on the chiral magnetism in this rare-earth system, knowledge on the CEF level scheme is indispensable. 
Although the energy levels have been estimated from the Schottky anomaly in specific heat 
and a possible CEF parameters are proposed by the analysis of the $M(H)$ curves~\cite{Yamashita12,Ito20}, the level scheme should be confirmed by inelastic neutron scattering (INS), which is the most direct method to observe the magnetic excitations. 

This paper is organized as follows. 
After describing the experimental procedure in Sec.~\ref{sec:Experiment}, we present in Sec.~\ref{sec:Results} the results of INS experiment and the analysis using the single-ion CEF model. 
The INS spectrum consists of well-defined magnetic excitations. The level scheme we conclude explains the intensity ratio between the peaks and the temperature ($T$) dependence fairly well. 
In Sec.~\ref{sec:DiscussionA}, we apply the single-ion CEF model to explain the magnetization curves, $T$-dependence of the lattice parameters of $c$ and $a$, and the Schottky anomaly in the magnetic specific heat. 
In Sec.~\ref{sec:DiscussionB}, we discuss the detailed properties of magnetic susceptibility that cannot be explained by the single-ion model only. It is necessary to consider that the exchange interaction between the CEF ground states is ferromagnetic, whereas the interaction involving the excited state is antiferromagnetic. We discuss this model as an orbital dependent exchange interaction, which is also associated with the $T$-dependence of the CEF excitation in the INS intensity.

\section{Experiment}
\label{sec:Experiment}
Single crystals of YbNi$_3$Al$_9$ were prepared by an Al-flux method following the procedure described in the literature~\cite{Ohara14}. 
INS experiment has been performed using the triple-axis thermal neutron spectrometer TOPAN installed at the research reactor JRR-3, Japan Atomic Energy Agency, Tokai, Japan. A monochromatic incident beam was obtained by using the 002 Bragg reflection of pyrolytic graphite (PG) crystals.  The energy of the scattered neutrons was analyzed using a PG-002 analyzer at a constant final energy $E_f$=13.5 meV. 
A PG filter was placed before the third collimator to cut the higher harmonic neutrons. A sapphire filter was inserted before the second collimator to reduce the background. The collimator setting was open-30'-30'-60'. 
The measurement was performed using a collection of small pieces of single crystals with a total mass of 5.59 g consisting of more than 1000 pieces, which was wrapped in an aluminum foil and was sealed in an aluminum cell with helium exchange gas. 
The sample was cooled by using a closed-cycle refrigerator.  
The scattering vector was fixed at $Q=1.94$ \AA$^{-1}$ so that no Bragg diffraction was superimposed at the elastic position.  

\section{Results and analysis}
\label{sec:Results}
\begin{figure}[t]
\begin{center}
\includegraphics[width=8cm]{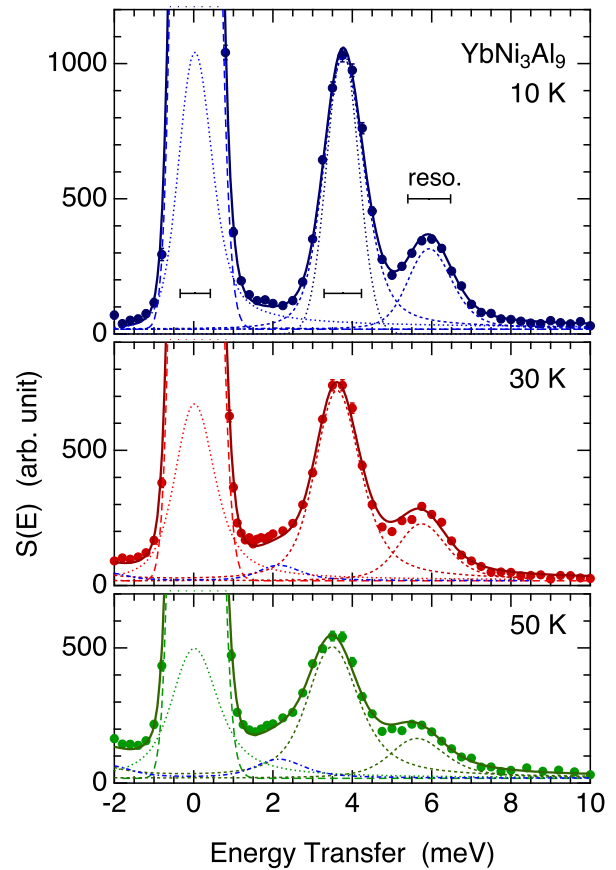}
\end{center}
\caption{\label{fig:INSspec}
Inelastic neutron scattering spectra of polycrystalline YbNi$_3$Al$_9$ at 10 K, 30 K, and 50 K. 
The solid lines are the fitting results in which the peak intensities, $f_{ij}$ in Eq. (\ref{eq:fij}), were fixed to the calculated values for the single-ion CEF model after taking the powder average. 
The dashed and the dotted lines are the profiles of the constituent peaks. The horizontal bars represent the instrumental energy resolution (full-width at half-maximum). 
}
\end{figure}

Figure \ref{fig:INSspec} shows the inelastic neutron scattering function $S(E)$, where $E$ represents the energy transfer $E=E_{i} - E_{f}$. The observed intensity $I(E)$ has been converted to $S(E)$ by correcting for the $E_i$ dependent monitor efficiency. 
The background intensity from the empty cell has been subtracted, which was measured under the same experimental condition. 
At the lowest temperature of 10 K, two inelastic peaks are clearly observed at 3.8 meV and at 5.9 meV. 
The widths are almost resolution limited, suggesting an almost dispersionless mode. 
At elevated temperatures of 30 K and 50 K, the intensities of the two peaks decreases and that around 2 meV increases, indicating that the former two peaks are the CEF excitations from the ground state to the excited states and the latter is due to the excitation from the 3.8 meV state to the 5.9 meV state. 
No significant signal was observed above 10 meV up to 22 meV. From the analysis of the specific heat of a nonmagnetic reference compound LuNi$_3$Al$_9$, the phonon Debye energy is estimated to be $\sim 16$ meV and the density of states is expected to extend up to approximately 50 meV. Therefore, the negligibly small signal above 10 meV shows that the phonon contribution is small and the inelastic signals below 10 meV can be ascribed to the magnetic excitations between the CEF levels. 
More detailed estimation of the phonon contribution to the intensity, using the data at 250 K, is described in the supplemental material (SM)~\cite{SM}.

Let us analyze the scattering function $S(E)$ using the single-ion CEF model. 
As proposed in Ref.~\onlinecite{Ito20}, we assume the following CEF Hamiltonian:
\begin{align}
\mathcal{H}_{\text{CEF}} &= 
B_{20}O_{20} + B_{40}O_{40} + B_{60}O_{60} + B_{66}O_{66} \nonumber \\
&= \begin{pmatrix}
a_{7} & 0 & 0 & 0 & 0 & 0 & a_{75} & 0 \\
0 & a_{5} & 0 & 0 & 0 & 0 & 0 & a_{75} \\
0 & 0 & a_{3} & 0 & 0 & 0 & 0 & 0 \\
0 & 0 & 0 & a_{1} & 0 & 0 & 0 & 0 \\
0 & 0 & 0 & 0 & a_{1} & 0 & 0 & 0 \\
0 & 0 & 0 & 0 & 0 & a_{3} & 0 & 0 \\
a_{75} & 0 & 0 & 0 & 0 & 0 & a_{5} & 0 \\
0 & a_{75} & 0 & 0 & 0 & 0 & 0 & a_{7} \\
\end{pmatrix} \;,
\label{eq:Hcef}
\end{align}
where $B_{lm}$ are the CEF parameters and $O_{lm}$ the Stevens operator equivalents~\cite{Stevens52,Hutchings64}. 
We neglect the $B_{43}$ and $B_{63}$ terms as explained later. 
The diagonal elements in (\ref{eq:Hcef}) are represented by 
$a_n = \langle \pm \frac{n}{2} | \mathcal{H}_{\text{CEF}} |\!\pm\!\frac{n}{2}\rangle $. 
The $B_{66}$ term gives the off-diagonal elements, which are represented by 
$a_{75} = \langle \pm \frac{7}{2} | \mathcal{H}_{\text{CEF}} | \!\mp\! \frac{5}{2} \rangle$.  
The CEF eigenstates are given by 
\begin{equation}
\begin{aligned}
|\varphi_{1 \pm}\rangle &= \alpha |\!\pm\! \frac{7}{2} \rangle - \beta  |\!\mp\! \frac{5}{2} \rangle \,, \\
|\varphi_{2 \pm}\rangle &= \beta |\!\pm\! \frac{7}{2} \rangle + \alpha  |\!\mp\! \frac{5}{2} \rangle \,,   \\
|\varphi_{3 \pm}\rangle &=  |\!\pm\! \frac{1}{2} \rangle \,,   \\
|\varphi_{4 \pm}\rangle &=  |\!\pm\! \frac{3}{2} \rangle \;.
\end{aligned} 
\end{equation}
From the analysis of $M(H)$, the ground state is expected to be either $|\varphi_1\rangle$ or $|\varphi_2\rangle$ with nearly equal coefficients $\alpha \approx \beta$~\cite{Ito20}. 
The magnitudes of the nonzero off-diagonal elements for the magnetic dipole moment are 
\begin{equation}
\begin{aligned}
|\langle \varphi_{1\pm} | J_{x,y} | \varphi_{2\mp} \rangle| &= \sqrt{7}|\alpha^2-\beta^2|/2  \,, \\
|\langle \varphi_{1\pm} | J_z | \varphi_{2\pm} \rangle| &= 6|\alpha\beta|  \,,  \\
|\langle \varphi_{1\pm} | J_{x,y} | \varphi_{4\mp} \rangle| &= \sqrt{3}|\beta|  \,, \\
|\langle \varphi_{2\pm} | J_{x,y} | \varphi_{4\mp} \rangle| &= \sqrt{3}|\alpha|  \,, \\
|\langle \varphi_{3\pm} | J_{x,y} | \varphi_{4\pm} \rangle| &= \sqrt{15}/2   \,.
\end{aligned} 
\end{equation}
A full description of the matrix elements is given in the SM~\cite{SM}. 
If we assume $\varphi_1$ to be the ground state, two excitations to $\varphi_2$ and $\varphi_4$ are allowed at the lowest temperature. 
Since $|\langle \varphi_{1\pm} | J_z | \varphi_{2\pm} \rangle|$ is larger than $|\langle \varphi_{1\pm} | J_{x,y} | \varphi_{4\mp} \rangle|$ when $\alpha \approx \beta$, the intensity ratio at 10 K implies that $\varepsilon_2 - \varepsilon_1 = 44$ K and $\varepsilon_4 - \varepsilon_1 = 69$ K. 
To know $\varepsilon_3$, we need to observe the transition between $\varphi_3$ and $\varphi_4$ at elevated temperatures. 

Since no clear peak corresponding to this transition is identified in the data at 30 K and 50 K, we estimate $\varepsilon_3$ from the Schottky anomaly in specific heat~\cite{Yamashita12}. 
By calculating $C_{\text{mag}}(T)$, we see that $\varepsilon_3$ should be located between 60 K and 80 K so that $C_{\text{mag}}(T)$ peaks at 20 K. 
If this is the case, $\varepsilon_3$ and $\varepsilon_4$ are almost degenerate, and the inelastic peak corresponding to this transition is expected to be less than 1 meV, which is hidden in the elastic background in the present experiment. 

\begin{figure}[t]
\begin{center}
\includegraphics[width=8.5cm]{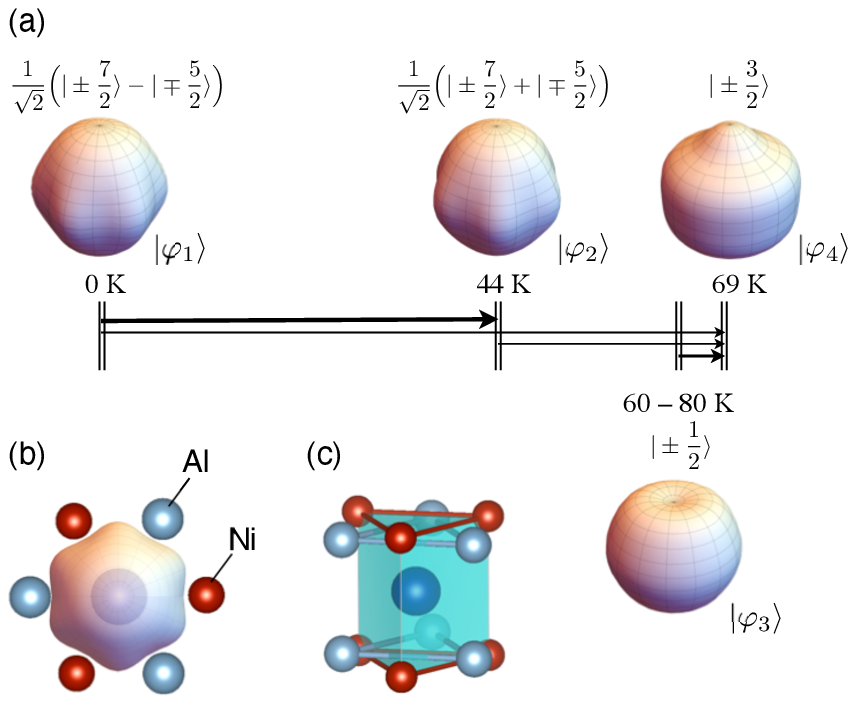}
\end{center}
\caption{(a) Crystal-field level scheme of YbNi$_3$Al$_9$. The energy level of the $|\pm \frac{1}{2} \rangle$ state remains uncertain, which should be between 60 K and 80 K. The arrows between the doublets represent the allowed magnetic transitions in INS. 
(b) Top view of the local environment of Yb and the $4f$ charge distribution of the $\varphi_1$ wave function. 
(c) Local structure around Yb constructed from the nearest neighbor Ni and Al atoms.  
}
\label{fig:LevelScheme}
\end{figure}

An ideal case of $\alpha=\beta=1/\sqrt{2}$ is obtained when $a_7=a_5$. 
In this case, $a_{75}=22$ K gives the splitting $\varepsilon_2 - \varepsilon_1 = 44$ K, which is equivalent to $B_{66}=0.0231$ K. 
By setting the unknown energy $\varepsilon_3-\varepsilon_1$, we obtain the other three parameters.
For example, when $\varepsilon_3-\varepsilon_1=64$ K, $B_{20}=-1.393$ K, $B_{40}=0.0128$ K, and $B_{60}=0.00128$ K are deduced, which give $a_{7}=a_{5}=-22.25$ K, $a_{3}=24.75$ K, and $a_{1}=19.75$ K. 
Figure~\ref{fig:LevelScheme}(a) shows the energy level scheme and the calculated charge distribution~\cite{Walter86}.

The solid lines in Fig.~\ref{fig:INSspec} are the fits with the following function:
\begin{align}
S(E) &=  \frac{CE}{1-e^{-E/k_{\text{B}}T}} \sum_{i,j} f_{ij} P(E;\Delta_{ij}, \Gamma) \label{eq:SE} \;,\\
f_{ij} &= \frac{2 e^{-\varepsilon_i/k_{\text{B}}T}}{Z} \frac{| \langle \varphi_i | J_{\perp} | \varphi_j \rangle |^2}{\varepsilon_j - \varepsilon_i} \label{eq:fij} \;,
\end{align}
where $P(E;\Delta_{ij},\Gamma)$ represents the Lorentzian spectral function peaked at $E=\Delta_{ij}$ with a half-width at half-maximum (HWHM) $\Gamma$, which is assumed to be independent of the CEF state. 
$f_{ij}$ is the theoretical intensity for the transition between $\varphi_i$ and $\varphi_j$.  
$J_{\perp}$ represents the magnetic dipole moment perpendicular to the scattering vector. 
The fitting has been carried out by using the calculated intensity for the ideal wave functions with $\alpha=\beta=1/\sqrt{2}$. 
Only the peak position $\Delta_{ij}=\varepsilon_j - \varepsilon_i$, the HWHM, and the scale factor $C$ were treated as fitting parameters. 
This means that the intensity of the three inelastic peaks corresponding to $(i,j)$=(1,2), (1,4), and (2,4) are fixed at the calculated values, which are obtained by taking the powder average. 
The intensity of the quasi-elastic peak centered at $E=0$ were treated as a free parameter to fit the tail of the elastic peak.  
The incoherent background at $E=0$ was also included in the fitting. 
The fitting was carried out by taking the convolution with the instrumental resolution. 
The resultant parameters are summarized in Table~\ref{tb:1}. 
The uncertainty values represent $3\sigma$ of the fit, where the experimental resolution was assumed to be exact.

\begin{table}[t]
\caption{Parameters obtained from the fitting. The number in the parenthesis represents the uncertainty in the last digits of the parameter. }
\label{tb:1}
\begin{tabular}{ccccc}
\hline
T (K) & $\Delta_{12}$ (meV) & $\Delta_{14}$ (meV) & $\Gamma$ (meV) & $C$ \\
\hline
10 K & 3.76(2) & 5.93(4) & 0.21(2) & 370(9) \\
30 K & 3.60(2) & 5.74(6) & 0.34(3) & 450(12) \\
50 K & 3.48(3) & 5.62(9) & 0.48(3) & 500(16) \\
\hline
\end{tabular}
\end{table}

As shown by the solid lines in Fig.~\ref{fig:INSspec}, the data are well explained by the calculated intensities based on the level scheme and the wave functions shown in Fig.~\ref{fig:LevelScheme}(a). 
The excitation energy decreases and the HWHM increases with increasing the temperature. These results can be attributed to the exchange interaction between the localized $f$-electrons and the itinerant conduction electrons~\cite{Becker77,Maekawa85,Otsuki05}. 
The slight increase in the scale factor $C$ at elevated temperatures shows that the single-ion CEF model is insufficient to explain the data. 
This point will be discussed in Sec.~\ref{sec:DiscussionB}. 
Experimentally, the $T$-dependence of the total INS intensity is consistent with that of the static magnetic susceptibility $\chi(T)$. 
Since $\chi(T)$ also deviates from the single-ion CEF calculation, it is natural that the $T$-dependence of the INS intensity also deviates from the single-ion CEF calculation. 
This will be explained by considering an exchange interaction in a mean-field approximation.

It is noted that the sequence of $\varphi_1$ and $\varphi_2$ cannot be determined in the present experiment and analysis.
It is associated with the sign of $B_{66}$ and the sequence in Fig.~\ref{fig:LevelScheme}(a) is reversed if we take a negative $B_{66}$, which does not affect the INS spectrum. 
The relation of the charge distribution of $\varphi_1$ illustrated in Fig.~\ref{fig:LevelScheme}(b) and the atomic positions of Ni and Al is also schematic; the relation between the CEF $x$-axis and the crystallographic $a$-axis is unknown and has an ambiguity of 30$^{\circ}$ rotation. 
To clarify these points, a precise measurement of the magnetic anisotropy in the $ab$-plane could provide useful information. 
More directly, nonresonant inelastic x-ray scattering may resolve the sign of $B_{66}$ as has been applied to determine the sign of $B_{44}$  in tetragonal CeCu$_2$Si$_2$ and CeCu$_2$Ge$_2$~\cite{Willers12,Rueff15}. 
Hard x-ray photoemission spectroscopy is also a candidate method~\cite{Aratani18}. 

Since the local symmetry of Yb is $C_3$ in a strict sense, the $B_{43}$ and $B_{63}$ terms should be included. 
This is because the almost equilateral-triangular-prism shaped environment of the Yb ion as illustrated in Fig.~\ref{fig:LevelScheme}(c) is actually not the perfect one. 
The upper and the lower triangles are different and twisted, which give rise to $B_{43}$ and $B_{63}$~\cite{Gladyshevskii93,Nakamura20}. 
If these terms exist, other off-diagonal elements appear in Eq.~(\ref{eq:Hcef}), $|M\rangle$ and $|M\pm 3\rangle$ states mix, and therefore all the three peaks should arise at 10 K. 
However, the additional peak is too small to be detected, which guarantees our assumption of the CEF Hamiltonian in Eq.~(\ref{eq:Hcef}).

\begin{figure}[t]
\begin{center}
\includegraphics[width=8cm]{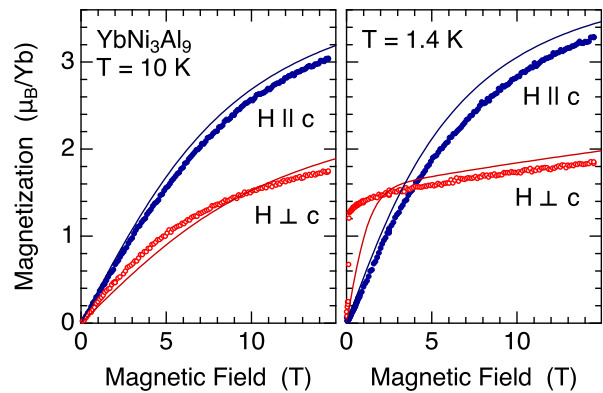}
\end{center}
\caption{
Magnetization curves of YbNi$_3$Al$_9$ at 10 K and 1.4 K for $H \parallel c$ and $H \perp c$. 
Solid lines are the calculations for the single-ion CEF model (see text). 
}
\label{fig:magnetization}
\end{figure}

\section{Discussions}
\label{sec:Discussion}
\subsection{Comparison with the bulk properties}
\label{sec:DiscussionA}
To show the validity of the above CEF model, we compare in Fig.~\ref{fig:magnetization} the calculated and measured $M(H)$ curves at 10 K and 1.4 K for $H \parallel c$ and $H \perp c$. 
The solid lines are the calculations for the single-ion CEF model for the ideal wave functions with $\alpha=\beta=1/\sqrt{2}$ and $\varepsilon_3 - \varepsilon_1=64$ K. 
Note that the calculation does not take into account the internal field due to the magnetic order and does not explain the low field region at 1.4 K. 
The helical order in the $ab$-plane is due to the larger diagonal element of $|\langle \varphi_{1\pm} | J_{x,y} | \varphi_{1\mp} \rangle| = 1.32$ than $|\langle \varphi_{1\pm} | J_{z} | \varphi_{1\pm} \rangle| = 0.5$ (see SM for the full description of the matrix elements)~\cite{SM}. 
The saturation moment of 1.5 $\mu_{\text{B}}$ for $H \perp c$ is also due to this in-plane diagonal elements. The gradual increase at high fields is caused by the mixing with the excited states. 
The large off-diagonal element $|\langle \varphi_{1\pm} | J_{z} | \varphi_{2\pm} \rangle| = 3$ gives rise to the Van Vleck type magnetization for $H \parallel c$, which becomes important at high fields and at high temperatures. This causes the overtaking in the $M(H)$ curve at around 4 T. 
It is remarked that the $c$-axis becomes the CEF easy axis at high fields and at high temperatures when the excited levels are populated, whereas the $ab$-plane becomes the easy plane at low fields and at low temperatures when only the ground doublet is populated. 
The consistency with the data are improved than in Ref.~\onlinecite{Ito20}.

\begin{figure}[t]
\begin{center}
\includegraphics[width=8cm]{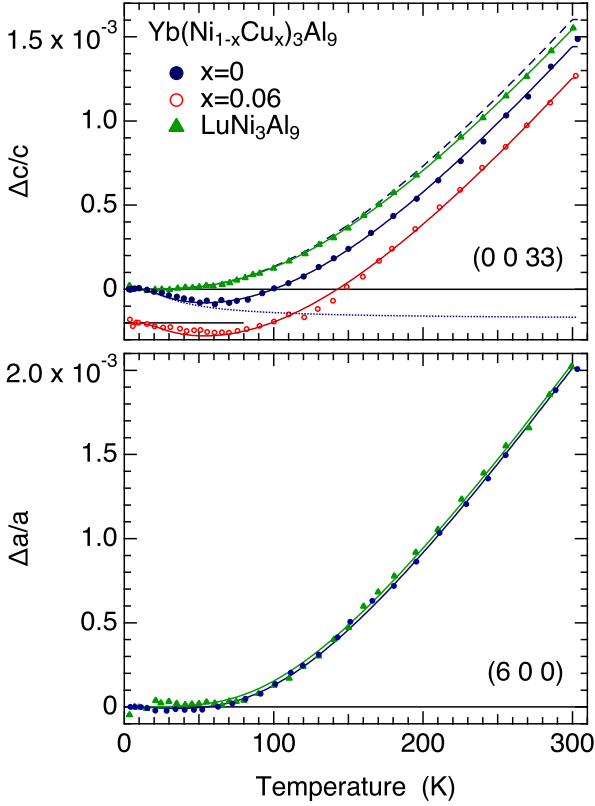}
\end{center}
\caption{
Temperature dependences of the relative change in the lattice parameters $c$ and $a$ obtained from the (0 0 33) and (6 0 0) reflection, respectively, by single-crystal x-ray diffraction.  
The solid lines are the fitting results as explained in the text. The dashed and dotted lines for $\Delta c/c$ represent the background lattice contribution and the magnetic contribution, respectively, for YbNi$_3$Al$_9$. 
}
\label{fig:Tdepdd}
\end{figure}

\begin{figure}[t]
\begin{center}
\includegraphics[width=8cm]{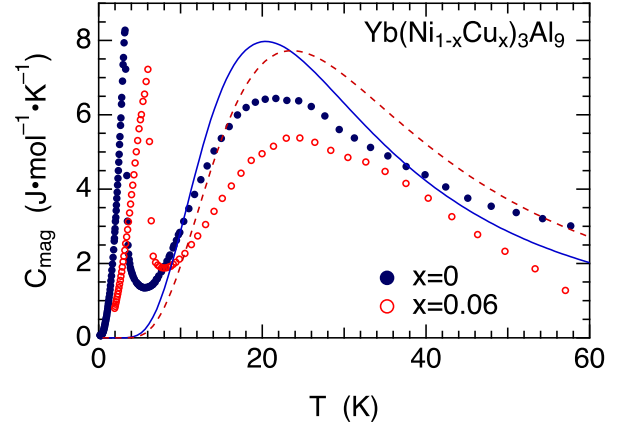}
\end{center}
\caption{
Comparison of the magnetic specific heat data at zero field with the calculation assuming a single-ion CEF model described in the text. 
}
\label{fig:spch}
\end{figure}

Another CEF effect is observed in the $T$-dependences of the lattice parameter shown in Fig.~\ref{fig:Tdepdd}, which were measured by single-crystal x-ray diffraction using a laboratory based Cu-$K_{\alpha}$ source.  
$\Delta c/c$ takes a minimum at around 50 K. 
Although this kind of minimum is often observed in valence fluctuating Yb compounds~\cite{Takeuchi10}, this case in YbNi$_3$Al$_9$ is more likely to be associated with the CEF effect since Yb is almost trivalent in the whole temperature range~\cite{Utsumi12}.  
The lattice parameter $c$ shrinks when the excited CEF states are more populated. 

The linear thermal expansion coefficient for a uniaxial pressure $p$ is expressed as $\alpha = (\partial L/\partial T)_p/L=\kappa \gamma C/V$, where $V=AL$ is the volume of the sample with the length $L$ and the cross-section $A$, $\kappa=-(\partial L/\partial p)_T/L$ the isothermal compressibility, $\gamma$ the Gr\"uneisen parameter, and $C$ the heat capacity. 
By treating $\kappa \gamma /V$ as a parameter respectively for the $c$ and $a$ axes, we can calculate $\Delta c/c$ and $\Delta a/a$ by modeling the heat capacity. The background curves of the lattice part were fit with the Debye and a constant density of states contributions, which are shown by the solid lines for LuNi$_3$Al$_9$ and the dashed line for YbNi$_3$Al$_9$. 
The $4f$-contribution to $\Delta c/c$ was treated as being proportional to the change in the thermal average of the ($3z^2 - r^2$)-type quadrupole moment, $\langle O_{20} \rangle$=$\{3\langle J_z^{\;2} \rangle - J(J+1)\}/2$; i.e., 
\begin{equation}
(\Delta c/c)_{4f}=K (\langle O_{20} \rangle_T - \langle O_{20} \rangle_{T=0}).
\end{equation}
Since the diagonal elements for the four CEF doublet states are $\langle O_{20} \rangle_1=\langle O_{20} \rangle_2 = 6$, $\langle O_{20} \rangle_3=-4.5$, $\langle O_{20} \rangle_4=-7.5$, and the second order Stevens factor for Yb$^{3+}$ is positive, the charge distribution of $f$ electrons of $\varphi_1$ and $\varphi_2$ are elongated along the $c$ axis and those of $\varphi_3$ and $\varphi_4$ are extended in the $ab$-plane. 
This leads to the decrease in $c$ when the excited levels are more populated at elevated temperatures. 
We obtained $K=2.91 \times 10^{-5}$ for YbNi$_3$Al$_9$. The result of the fit is shown by the solid and dotted lines in Fig.~\ref{fig:Tdepdd}. 
In contrast to $\Delta c/c$, $\Delta a/a$ does not exhibit such a minimum. 
This is explained by the fact that the ($x^2-y^2$)-type quadrupole moment, $O_{22}=\sqrt{3}(J_x^{\;2} - J_y^{\;2})/2$, and the $xy$-type quadrupole moment, $O_{xy}=\sqrt{3}(J_x J_y + J_y J_x)/2$, do not possess any diagonal element.

The calculated $C_{\text{mag}}(T)$ for the above level scheme with $\varepsilon_3=64$ K is compared with the data in Fig.~\ref{fig:spch}. 
The smaller peak height of the data is probably due to the Kondo effect~\cite{Shimizu96}.  
In a Cu substituted system with $x=0.06$, the peak of $C_{\text{mag}}(T)$ shifts to the higher temperature and the magnetization decreases~\cite{Ohara14,Ninomiya18}. 
These results show that the total energy splitting increases and that $\alpha$ and $\beta$ deviate from $1/\sqrt{2}$. 
For example, if we assume $B_{20}=-1.5$ K, $B_{40}=0.022$ K, $B_{60}=0.0022$ K, and $B_{66}=0.024$ K, the CEF level scheme becomes $\varepsilon_1$=0 K, $\varepsilon_2$=48.4 K, $\varepsilon_3$=72.2 K, and $\varepsilon_4$=86.2 K, 
%0 -- 51 -- 89 -- 106 K, 
where $\alpha=0.578$ and $\beta=0.816$. 
As shown by the dashed line in Fig.~\ref{fig:spch}, the tendency of the change in $C_{\text{mag}}(T)$ for $x=0.06$ can partly be explained. 
The decreasing tendency in $M(H)$ at high fields by the Ni substitution is also partly explained as shown in the SM~\cite{SM}.  
In any case, the CEF level scheme for $x=0.06$ needs to be studied in more detail. 
It is also necessary to take into account the stronger exchange interaction for $x=0.06$ with higher $T_{\text{N}}$ (=6.5 K) and $H_{\text{C}}$ (=1.0 T) than those for $x=0$. 
The solid line for $\Delta c/c$ of Yb(Ni$_{0.94}$Cu$_{0.06}$)$_3$Al$_9$ in Fig.~\ref{fig:Tdepdd} is the same as that for YbNi$_3$Al$_9$. 
This is because the $T$-dependence of $\Delta c/c$ is not as sensitive to this level scheme as $C_{\text{mag}}(T)$.

The almost ideal CEF states with $\alpha=\beta=1/\sqrt{2}$ in YbNi$_3$Al$_9$ is of course considered to be accidental. 
First, the negative $B_{20}$ prefers the $|\pm 7/2\rangle$ state to be the ground state, which makes the $c$ axis the easy axis.  Second, the $B_{40}$ and $B_{60}$ terms lower the energy level of the $|\pm 5/2\rangle$ state and accidentally make them almost degenerate. 
Third, the off-diagonal element due to the $B_{66}$ term mixes these states and lifts the degeneracy, and thereby allows the in-plane $J_{x,y}$ moment to have a large diagonal element. 
The ground doublet can therefore be treated as an $S=1/2$ spin system with an anisotropic $g$-factor, where $g_{xy}=3.02$ and $g_z=1.14$. 

\subsection{Orbital dependent exchange interaction}
\label{sec:DiscussionB}
\begin{figure}[t]
\begin{center}
\includegraphics[width=8cm]{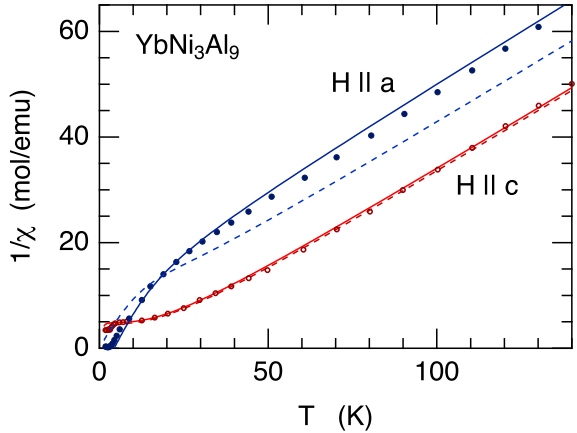}
\end{center}
\caption{
Temperature dependences of magnetic susceptibility measured at 0.1 T. The data are taken from the literature~\cite{Yamashita12}.  
The dashed lines are the calculations for the single-ion CEF model. 
The solid lines consider an orbital-dependent exchange interaction in a mean-field approximation. 
}
\label{fig:magsus}
\end{figure}

Although the CEF level scheme deduced from the INS data analysis generally explains the bulk properties well, the single-ion CEF model is insufficient to explain the $T$-dependence of the INS intensity in detail as shown by the $T$-dependent scale factor, which should in principle be a constant. 
This is a natural consequence because the single-ion CEF model itself is insufficient to explain the static magnetic susceptibility in detail, especially at low temperatures where the exchange interaction plays an important role. 
In Fig.~\ref{fig:magsus}, we compare the inverse magnetic susceptibility with the calculation for the single-ion model with $\varepsilon_3=64$ K and $\alpha=\beta=1/\sqrt{2}$ as assumed in this paper. 
For $H \parallel c$, the data are well reproduced by the single-ion model in the whole temperature range. 
For $H \parallel a$, the data at high temperatures above 50 K, where the CEF excited levels are populated, can be explained by vertically shifting the calculated curve, i.e., by considering a uniform antiferromagnetic exchange interaction in a normal mean-field approximation. 
However, with decreasing temperature, especially below 30 K where the population of the ground state increases, the downward curvature of the $1/\chi$ data is enhanced, suggesting an increase in the ferromagnetic interaction. 
These features suggest that the exchange interaction is not uniform and is dependent on the CEF states. It is necessary to consider an orbital dependent exchange interaction~\cite{Ito20}. 

We assume that the molecular fields originating from different CEF states have different effects~\cite{Aoki80}.  
The total magnetic moment $\langle \mu \rangle$ in a weak magnetic field $H$ is expressed as the sum of the diagonal (Curie) and the off-diagonal (Van Vleck) contributions 
\begin{equation}
\langle \mu \rangle = \sum_i \langle \mu_i \rangle + \sum_{i,j} \langle \mu_{[ij]} \rangle \;,
\end{equation}
where $\langle \mu_i \rangle$ represents the diagonal moment of the CEF state $|\varphi_i\rangle$ and $\langle \mu_{[ij]} \rangle$ the Van Vleck moment induced between the states $|\varphi_i\rangle$ and $|\varphi_j\rangle$. 
These moments are expressed as
\begin{align}
\langle \mu_i \rangle &= \chi_i^{(0)} \bigl( H + \sum_j \lambda_{ij} \langle \mu_j \rangle + \sum_{j,k} \lambda_{i[jk]} \langle \mu_{[jk]} \rangle \bigr) \,, \\
\langle \mu_{[ij]} \rangle &=  \chi_{[ij]}^{(0)} \bigl( H + \sum_k \lambda_{k[ij]} \langle \mu_k \rangle + \sum_{k,l} \lambda_{[ij][kl]} \langle \mu_{[kl]} \rangle \bigr) \,,
\end{align}
where $\chi_i^{(0)}$ is the Curie susceptibility of the CEF state $|\varphi_i\rangle$ and $\chi_{[ij]}^{(0)}$ the Van Vleck susceptibility between the states $|\varphi_i\rangle$ and $|\varphi_j\rangle$ in the single-ion calculation. $\lambda_{ij}$, $\lambda_{i[jk]}$, and $\lambda_{[ij][kl]}$ are the mean-field exchange parameter between $\langle \mu_i \rangle$ and $\langle \mu_j \rangle$, $\langle \mu_i \rangle$ and $\langle \mu_{[jk]} \rangle$, and between $\langle \mu_{[ij]} \rangle$ and $\langle \mu_{[jk]} \rangle$, respectively. 
In the present case with four CEF levels, we have ten exchange parameters (four Curie and six Van Vleck) in total. 
Then, to reduce the number of parameters for simplicity, we approximate the above equations as the following: 
\begin{align}
\langle \mu \rangle &=  \langle \mu_{\text{c}} \rangle +  \langle \mu_{\text{v}} \rangle \,, \\
\langle \mu_{\text{c}} \rangle &= \chi_{\text{c}}^{(0)} \big( H + \lambda_{\text{cc}} \langle \mu_{\text{c}} \rangle + \lambda_{\text{cv}} \langle \mu_{\text{v}} \rangle \bigr) \,, \\
\langle \mu_{\text{v}} \rangle &= \chi_{\text{v}}^{(0)} \big( H + \lambda_{\text{cv}} \langle \mu_{\text{c}} \rangle + \lambda_{\text{vv}} \langle \mu_{\text{v}} \rangle \bigr) \,,
\end{align}  
where $\langle \mu_{\text{c}} \rangle$ and $\langle \mu_{\text{v}} \rangle$ are the Curie and the Van Vleck moments, respectively. 
When $\lambda_{\text{cc}}=\lambda_{\text{cv}}=\lambda_{\text{vv}}$ (uniform exchange), we have a normal mean-field model. 

The solid lines in Fig.~\ref{fig:magsus} show the calculated $1/\chi$ curves obtained by assuming $\lambda_{\text{cc}}^{(a)}=\lambda_{\text{cc}}^{(c)}=5$, $\lambda_{\text{cv}}^{(a)}=\lambda_{\text{cv}}^{(c)}=-3$, $\lambda_{\text{vv}}^{(a)}=-20$, and $\lambda_{\text{vv}}^{(c)}=0$. Note that these parameters are renewed from those in Ref.~\onlinecite{Ito20} by using the renewed CEF level scheme in this paper. 
By introducing a ferromagnetic exchange between the Curie moments and an antiferromagnetic exchange between the Van Vleck  moments, the increase in the ferromagnetic response in $\chi_a$ for $H \parallel a$ below 30 K is well reproduced. 
This is due to the increasing contribution from the Curie-term susceptibility of $\langle \varphi_{1\pm} | J_x | \varphi_{1\mp} \rangle$ with ferromagnetic interaction. 
The ferromagnetic interaction of the ground state is also a key parameter in determining the long range ordering temperature~\cite{Shinozaki16}.  
At high temperatures where the excited levels are populated, the antiferromagnetic interaction becomes more important. 

\begin{figure}[t]
\begin{center}
\includegraphics[width=8.5cm]{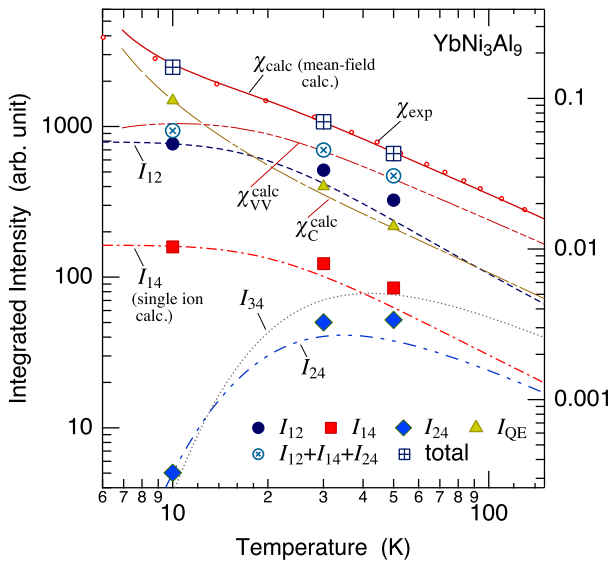}
\end{center}
\caption{
Temperature dependences of the integrated intensities of the 1--2 ($I_{12}$), 1--4 ($I_{14}$), 2--4 ($I_{24}$), and 3--4 ($I_{34}$) inelastic transitions between the CEF levels (left axis). The experimental data at 10 K, 30 K, and 50 K, are represented by the marks. The calculated intensities for the single-ion CEF model are represented by the dashed ($I_{12}=Cf_{12}$), dot-dashed ($I_{14}=Cf_{14}$), double-dot-dashed ($I_{24}=Cf_{24}$), and dotted ($I_{34}=Cf_{34}$) lines. The scale factor $C$ for the calculation is fixed to the value at 10 K. 
The intensities of the quasi-elastic (QE) peak and the total intensities are also shown by the marks. 
The open circles indicated by $\chi_{\text{exp}}$  (right axis) are the temperature dependence of the static magnetic susceptibility $(2\chi_a+\chi_c)/3$ (powder average), obtained from the data in the literature~\cite{Yamashita12}.  The solid line indicated by $\chi_{\text{calc}}$ represents the calculation for a mean-field model described in the text. The constituent contributions from the Van Vleck and the Curie terms are shown by the lines indicated by  $\chi_{\text{VV}}^{\text{calc}}$ and $\chi_{\text{C}}^{\text{calc}}$, respectively. 
The scales of the left and the right axes are proportional to each other. 
}
\label{fig:INSInt}
\end{figure}

Let us finally come back to the INS intensity. In Fig.~\ref{fig:INSInt}, we compare the experimentally obtained INS intensities at 10 K, 30 K, and 50 K with the calculated intensities for the single-ion model. 
The intensity is expressed by $I_{ij}=Cf_{ij}$ in Eq.~(\ref{eq:SE}); as for the experimental intensity, $C$ includes the $T$-dependence as summarized in Table~\ref{tb:1}, whereas for the calculation $C$ is a constant. The experimental intensities deviate from the calculations at high temperatures. 
On the other hand, if we compare the the total magnetic intensity, including the quasi-elastic one, with the static magnetic susceptibility after taking the powder average ($\chi_{\text{exp}}=(2\chi_a + \chi_c)/3$), they are consistent. 
This shows that the total intensity is properly estimated. 
Therefore, it is necessary to take into account the aforementioned orbital dependent exchange interaction to calculate the INS intensity. 
In other words, the experimental INS intensities of $I_{12}$, $I_{14}$, and $I_{24}$ include consequences of such exchange interactions. 
In Fig.~\ref{fig:INSInt}, the experimental $I_{12}+I_{14}+I_{24}$ (not including $I_{34}$) is compared with $\chi_{\text{VV}}^{\text{calc}}$, the Van Vleck part of $\chi_{\text{calc}}$ ($\propto I_{12}+I_{14}+I_{24}+I_{34}$). Also, the experimental $I_{\text{QE}}$ (including $I_{34}$) is compared with $\chi_{\text{C}}^{\text{calc}}$, the Curie part of $\chi_{\text{calc}}$ ($\propto I_{11}+I_{22}+I_{33}+I_{44}$). 
The $T$-dependences of these intensities are reasonably well reproduced by the mean-field calculation. 

The $T$-dependence of the intensity of the CEF excitation is little affected by the exchange interaction between the localized $f$-electrons and the conduction electrons when the $f$-electrons are well localized, as is the case in YbNi$_3$Al$_9$. 
Although the peak width is broadened and the effective excitation energy is weakly modified by the exchange interaction, the $T$-dependence of the peak intensity changes little from that of the single-ion model without exchange interaction~\cite{Becker77,Maekawa85}. 
The deviation of the intensity from the single-ion calculation as shown in Fig.~\ref{fig:INSInt}, therefore, cannot be explained by the Kondo effect, but should be attributed to the inter-ionic exchange interaction between localized moments. 
To confirm the orbital dependent exchange interaction proposed here, more precise measurement of the $T$-dependence of the intensities is necessary, especially the measurement of the quasi-elastic peak intensity by using cold neutrons, and thereby analyze separately the Curie term and the Van Vleck term susceptibilities. 

\section{Summary}
\label{sec:Summary}
In summary, we have performed an INS experiment on a uniaxial chiral helimagnet YbNi$_3$Al$_9$ and determined the CEF level scheme and the wave functions. The ground and the first excited doublets are located at 0 K and 44 K, respectively, and are described approximately by $(|\pm 7/2 \rangle - |\mp 5/2 \rangle)/\sqrt{2}$ or $(|\pm 7/2 \rangle + |\mp 5/2 \rangle)/\sqrt{2}$, which consistently explain the easy-plane anisotropy of the ground state and the helimagnetic ordering in the $ab$-plane. 
It also explains the easy axis anisotropy along the $c$ axis at high magnetic fields and high temperatures where the excited levels are involved. 
The $|\pm 3/2\rangle$ state is at 69 K and the $|\pm 1/2\rangle$ state is expected to be between 60 K and 80 K. 
We also analyzed the static magnetic susceptibility by taking into account the orbital dependent exchange interaction. 
With respect to the magnetic susceptibility in the $ab$-plane, the ferromagnetic interaction between the diagonal magnetic moments of the ground doublet plays an important role at low temperatures, whereas at high temperatures the antiferromagnetic interaction between the off-diagonal moments becomes important, which is also reflected in the INS intensity. 

\section*{Acknowledgments}
We thank Y. Kawamoto for the help in the INS experiment. 
This work was supported by JSPS KAKENHI Grants No. JP20H01854 and No. JP21K03467. The INS experiment was performed under the Joint-Use Research Program for Neutron Scattering, Institute for Solid State Physics (ISSP), The University of Tokyo, at the Research Reactor JRR-3, JAEA (Proposal No. 21516) and the 6G-TOPAN IRT program (Proposal No. 22402). We gratefully acknowledge the strong support of the Center of Neutron Science for Advanced Materials, Institute for Materials Research, Tohoku University. 
The x-ray diffraction experiment was performed by using a diffractometer at N-BARD, Hiroshima University. 
MT is supported by JST, the establishment of university fellowships towards the creation of science technology innovation, Grant No. JPMJFS2129.

%\appendix
%\section{}

%\begin{acknowledgments}
%\end{acknowledgments}

\bibliography{YbNi3Al9CEF3}

\clearpage
\begin{widetext}

\begin{center}
\textbf{\LARGE{Supplemental Material}}
\end{center}
\vspace{2mm}

\begin{center}
\textbf{\Large{Crystal field excitation in a chiral helimagnet YbNi$_3$Al$_9$}} \\
\vspace{4mm}
M. Tsukagoshi, S. Kishida, K. Kurauchi, D. Ito, K. Kubo, T. Matsumura, Y. Ikeda, S. Nakamura, and S. Ohara
\end{center}
\vspace{10mm}

\end{widetext}

\renewcommand{\topfraction}{1.0}
\renewcommand{\bottomfraction}{1.0}
\renewcommand{\dbltopfraction}{1.0}
\renewcommand{\textfraction}{0.01}
\renewcommand{\floatpagefraction}{1.0}
\renewcommand{\dblfloatpagefraction}{1.0}
\setcounter{topnumber}{5}
\setcounter{bottomnumber}{5}
\setcounter{totalnumber}{10}

\renewcommand{\theequation}{S\arabic{equation}}
\renewcommand{\thefigure}{S\arabic{figure}}
\renewcommand{\thetable}{S-\Roman{table}}
\setcounter{section}{19}
\setcounter{equation}{0}
\setcounter{figure}{0}
\setcounter{page}{1}

\subsection*{Dipole matrix elements for the CEF eigenstates}
We summarize here the full matrix elements of the magnetic dipole operators, $J_x$, $J_y$, and $J_z$, for the CEF eigenstates $|\varphi_{i\pm}\rangle$ $(i=1, 2, 3, 4)$. As explained in the main text, the eigenfunctions of 
\[
\mathcal{H}_{\text{CEF}}=B_{20}O_{20} + B_{40}O_{40} + B_{60}O_{60} + B_{66}O_{66}
\]
are written as
\begin{align}
|\varphi_{1 \pm}\rangle &= \alpha |\!\pm\! \frac{7}{2} \rangle - \beta  |\!\mp\! \frac{5}{2} \rangle \,, \\
|\varphi_{2 \pm}\rangle &= \beta |\!\pm\! \frac{7}{2} \rangle + \alpha  |\!\mp\! \frac{5}{2} \rangle \,,   \\
|\varphi_{3 \pm}\rangle &=  |\!\pm\! \frac{1}{2} \rangle \,,   \\
|\varphi_{4 \pm}\rangle &=  |\!\pm\! \frac{3}{2} \rangle \;.
\end{align} 
The matrices are 
\begin{widetext}
\begin{equation}
J_x = 
%\left( \begin{array}{cccccccc}
\bordermatrix{ & |\varphi_{1+}\rangle &  |\varphi_{1-}\rangle &  |\varphi_{2+}\rangle &  |\varphi_{2-}\rangle &   |\varphi_{3+}\rangle &  |\varphi_{3-}\rangle &  |\varphi_{4+}\rangle &  |\varphi_{4-}\rangle \cr
\langle \varphi_{1+}| & 0 & -\sqrt{7} \alpha \beta & 0 & \frac{\sqrt{7}(\alpha^2-\beta^2)}{2} & 0 & 0 & 0 & -\sqrt{3} \beta \cr
\langle \varphi_{1-}| & -\sqrt{7} \alpha \beta & 0 & \frac{\sqrt{7}(\alpha^2-\beta^2)}{2} & 0 & 0 & 0 & -\sqrt{3} \beta & 0 \cr
\langle \varphi_{2+}| & 0 &\frac{\sqrt{7}(\alpha^2-\beta^2)}{2} & 0 & \sqrt{7} \alpha \beta & 0 & 0 & 0 & \sqrt{3} \alpha \cr
\langle \varphi_{2-}| &\frac{\sqrt{7}(\alpha^2-\beta^2)}{2} & 0 & \sqrt{7}\alpha \beta & 0 & 0 & 0 & \sqrt{3} \alpha & 0 \cr
\langle \varphi_{3+}| & 0 & 0 & 0 & 0 & 0 & 2 & \frac{\sqrt{15}}{2} & 0 \cr
\langle \varphi_{3-}| & 0 & 0 & 0 & 0 & 2 & 0 & 0 & \frac{\sqrt{15}}{2} \cr
\langle \varphi_{4+}| & 0 & -\sqrt{3} \beta & 0 & \sqrt{3} \alpha & \frac{\sqrt{15}}{2} & 0 & 0 & 0 \cr
\langle \varphi_{4-}| & -\sqrt{3} \beta & 0 & \sqrt{3} \alpha & 0 & 0 & \frac{\sqrt{15}}{2} & 0 & 0 \cr
} \,, 
%\end{array} \right)
\end{equation}

\begin{equation}
J_y = 
%\left( \begin{array}{cccccccc}
\bordermatrix{ & |\varphi_{1+}\rangle &  |\varphi_{1-}\rangle &  |\varphi_{2+}\rangle &  |\varphi_{2-}\rangle &   |\varphi_{3+}\rangle &  |\varphi_{3-}\rangle &  |\varphi_{4+}\rangle &  |\varphi_{4-}\rangle \cr
\langle \varphi_{1+}| &  0 & i \sqrt{7} \alpha \beta & 0 & -i \frac{\sqrt{7}(\alpha^2- \beta^2)}{2}  & 0 & 0 & 0 & -i \sqrt{3} \beta \cr
\langle \varphi_{1-}| &  -i \sqrt{7} \alpha \beta & 0 & i\frac{\sqrt{7}(\alpha^2- \beta^2)}{2}   & 0 & 0 & 0 & i \sqrt{3} \beta& 0 \cr
\langle \varphi_{2+}| &  0 & -i \frac{\sqrt{7}(\alpha^2- \beta^2)}{2} & 0 & -i\sqrt{7} \alpha \beta & 0 & 0 & 0 & i \sqrt{3} \alpha \cr
\langle \varphi_{2-}| &  i \frac{\sqrt{7}(\alpha^2- \beta^2)}{2} & 0 & i\sqrt{7} \alpha \beta & 0 & 0 & 0 & -i \sqrt{3} \alpha & 0 \cr
\langle \varphi_{3+}| &  0 & 0 & 0 & 0 & 0 & -2 i & \frac{i \sqrt{15}}{2} & 0 \cr
\langle \varphi_{3-}| & 0 & 0 & 0 & 0 & 2 i & 0 & 0 & -\frac{i \sqrt{15}}{2} \cr
\langle \varphi_{4+}| &  0 & -i \sqrt{3} \beta & 0 & i \sqrt{3} \alpha & -\frac{i \sqrt{15}}{2} & 0 & 0 & 0 \cr
\langle \varphi_{4-}| &  i \sqrt{3} \beta & 0 & -i \sqrt{3} \alpha & 0 & 0 & \frac{i \sqrt{15}}{2} & 0 & 0 \cr
} \,, 
%\end{array} \right)
\end{equation}

\begin{equation}
J_z = 
%\left(\begin{array}{cccccccc} 
\bordermatrix{ & |\varphi_{1+}\rangle &  |\varphi_{1-}\rangle &  |\varphi_{2+}\rangle &  |\varphi_{2-}\rangle &   |\varphi_{3+}\rangle &  |\varphi_{3-}\rangle &  |\varphi_{4+}\rangle &  |\varphi_{4-}\rangle \cr
\langle \varphi_{1+}| & \frac{7 \alpha^2 - 5 \beta^2}{2} & 0 & 6 \alpha \beta & 0 & 0 & 0 & 0 & 0 \cr
\langle \varphi_{1-}| & 0 & -\frac{7 \alpha^2 - 5 \beta^2}{2} & 0 & -6 \alpha \beta & 0 & 0 & 0 & 0 \cr
\langle \varphi_{2+}| &  6 \alpha \beta & 0 & \frac{7 \beta^2 - 5 \alpha^2}{2} & 0 & 0 & 0 & 0 & 0 \cr
\langle \varphi_{2-}| & 0 & -6 \alpha \beta & 0 & - \frac{ 7 \beta^2 - 5 \alpha^2 }{2} & 0 & 0 & 0 & 0 \cr
\langle \varphi_{3+}|  & 0 & 0 & 0 & 0 & \frac{1}{2} & 0 & 0 & 0 \cr
\langle \varphi_{3-}|  & 0 & 0 & 0 & 0 & 0 & -\frac{1}{2} & 0 & 0 \cr
\langle \varphi_{4+}| & 0 & 0 & 0 & 0 & 0 & 0 & \frac{3}{2} & 0 \cr
\langle \varphi_{4-}| & 0 & 0 & 0 & 0 & 0 & 0 & 0 & -\frac{3}{2} \cr
} \,.
%\end{array} \right)
\end{equation}

\end{widetext}

\clearpage
\subsection*{Magnetization of Yb(Ni$_{\bm{1-x}}$Cu$_{\bm{x}}$)$_{\bm{3}}$Al$_{\bm{9}}$ for $\bm{x=0.06}$}
\begin{figure}[b]
\begin{center}
\includegraphics[width=8cm]{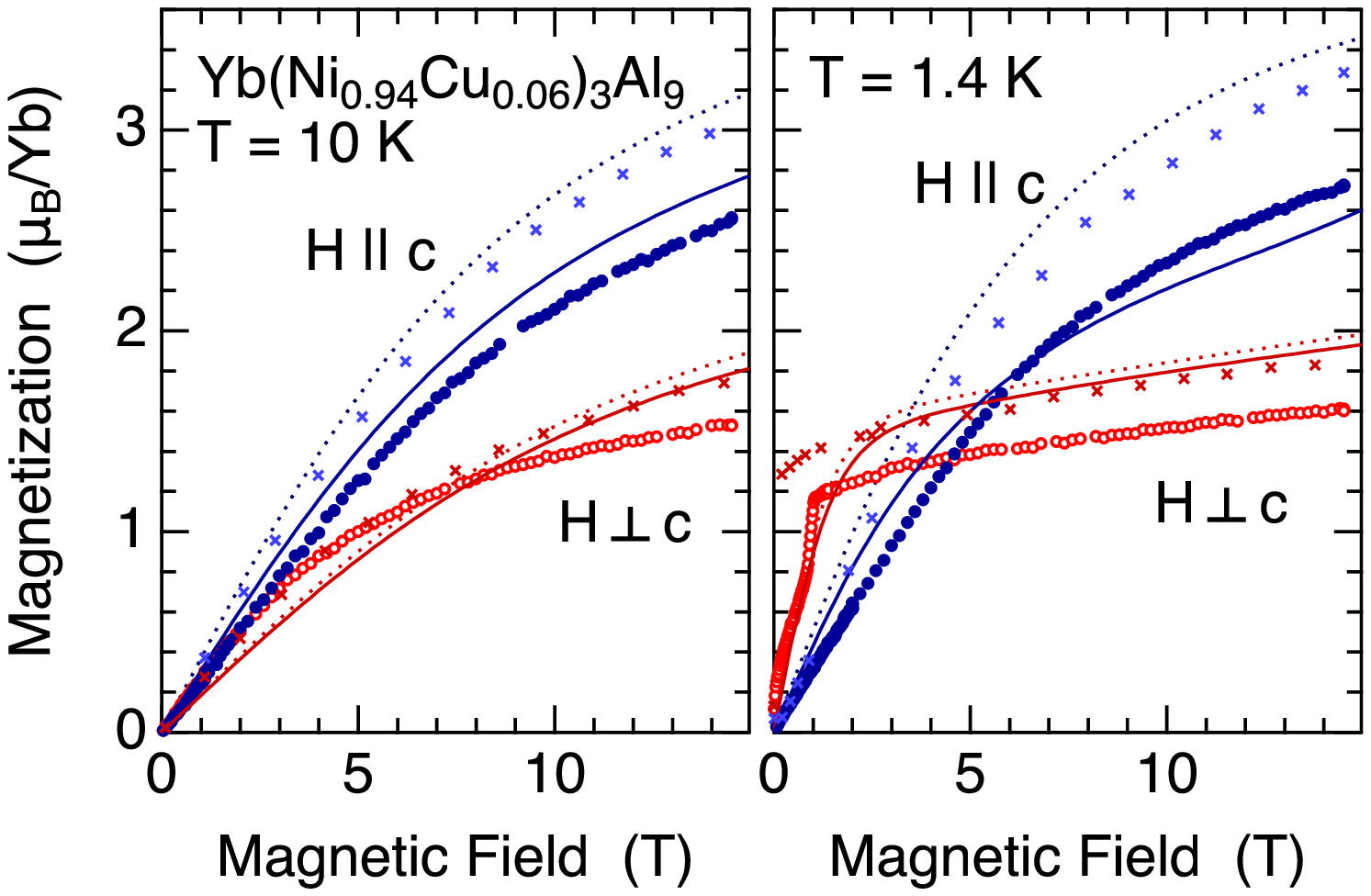}
\end{center}
\caption{
Magnetization curves of Yb(Ni$_{1-x}$Cu$_x$)$_3$Al$_9$ for $x=0.06$ at 10 K and 1.4 K for $H \parallel c$ and $H \perp c$. 
Solid lines are the calculations for the single-ion CEF model. Cross marks and the dotted lines are the data and the calculations, respectively, for $x=0$ shown in Fig.~\ref{fig:magnetization} in the main text. 
}
\label{fig:mag006SM}
\end{figure}

Figure \ref{fig:mag006SM} shows the magnetization curves of Yb(Ni$_{1-x}$Cu$_x$)$_3$Al$_9$ for $x=0.06$ at 10 K and 1.4 K for $H \parallel c$ and $H \perp c$. The data for $x=0$, the same as those in Fig.~\ref{fig:magnetization}, are also shown by the cross marks. 
These data show that $M(H)$ decreases by the Ni substitution, indicating that the CEF energy splitting increases. 
The decrease in $M(H)$ for $H \perp c$ implies that the coefficients $\alpha$ and $\beta$ deviate from $1/\sqrt{2}$. 

As described in the main text to discuss the magnetic specific heat $C_{\text{mag}}(T)$ for $x=0.06$, we assume $B_{20}=-1.5$ K, $B_{40}=0.022$ K, $B_{60}=0.0022$ K, and $B_{66}=0.024$ K. 
Then, we obtain energy levels of $\varepsilon_1=0$ K, $\varepsilon_2=48.4$ K, $\varepsilon_3=72.2$ K, and $\varepsilon_4=86.2$ K. 
The coefficients are $\alpha=0.578$ and $\beta=0.816$. 
The calculated $M(H)$ curves for the single-ion model are shown by the solid lines in Fig.~\ref{fig:mag006SM}. 
The tendency of the decrease in $M(H)$ is partly explained.  However, the detailed structures in the $M(H)$ curves are not reproduced well, especially the large decrease in $M(H)$ at high fields for $H \perp c$. 
In Yb(Ni$_{0.94}$Cu$_{0.06}$)$_3$Al$_9$ with a higher ordering temperature (=6.5 K) and a higher critical field for $H \perp c$ (=1.0 T), the antiferromagnetic exchange interaction is expected to be larger than that of YbNi$_3$Al$_9$. 
Therefore, it is more difficult to reproduce the $M(H)$ curves by the single-ion calculation than it is for YbNi$_3$Al$_9$. 
It is necessary to take into account the exchange interaction, which is a future work to be studied.

\newpage
\subsection*{Phonon contribution to the INS intensity}
\begin{figure}[b]
\begin{center}
\includegraphics[width=8cm]{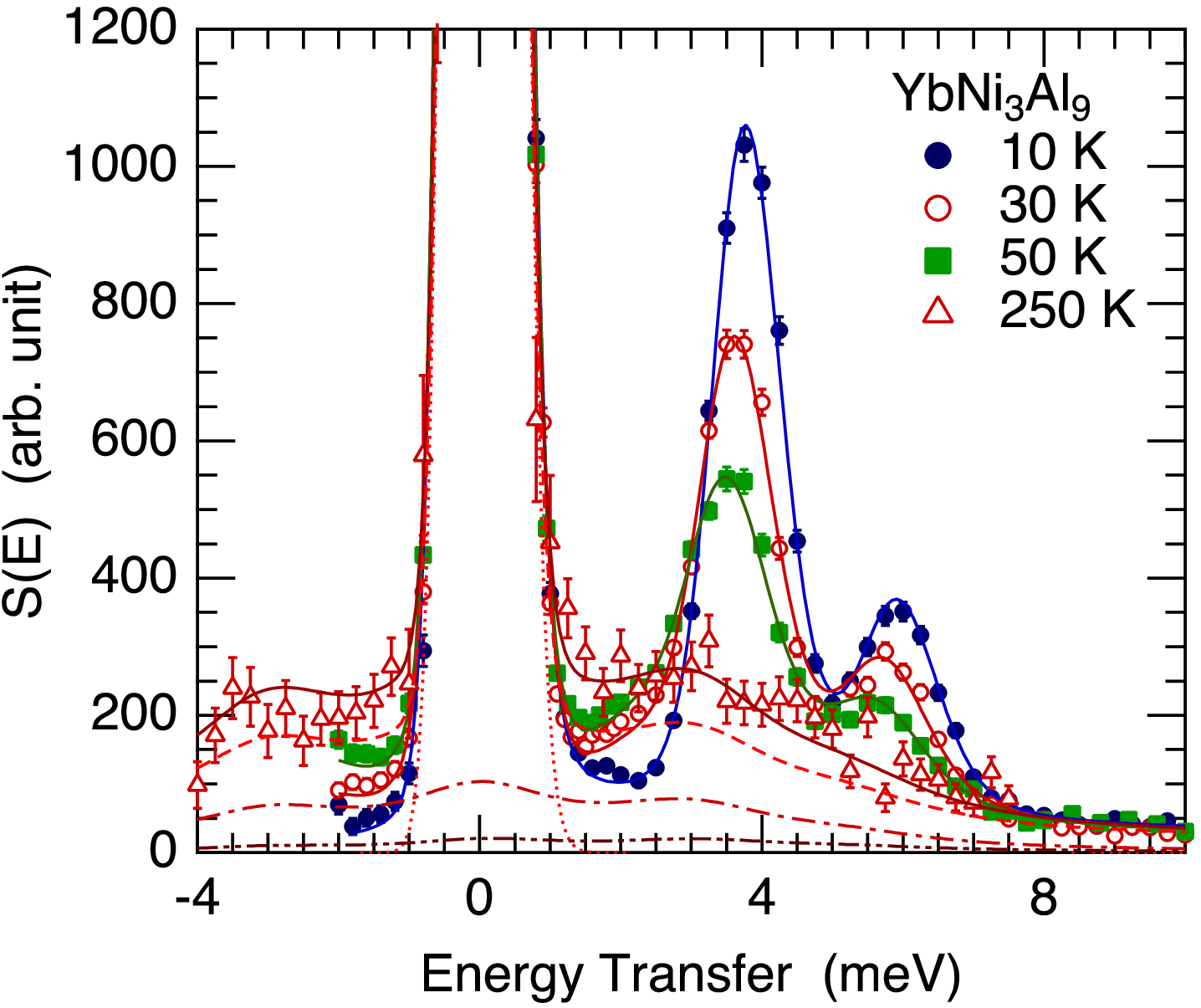}
\end{center}
\caption{
Inelastic neutron scattering spectra of YbNi$_3$Al$_9$ at 10 K, 30 K, 50 K (the same as those in Fig.~\ref{fig:INSspec} in the main text), and at 250 K. The solid lines are the results of fitting as described in the text. The dashed line shows the magnetic scattering expected from the single-ion CEF model by using the same scale factor at 10 K. The dot-dashed line shows the difference between the solid and the dashed lines, i.e., an estimated phonon contribution at 250 K.  
The double-dot-dashed line shows the expected phonon contribution at 50 K estimated by considering the Bose population factor. 
}
\label{fig:INSspecSM}
\end{figure}

The INS spectrum of YbNi$_3$Al$_9$ measured at 250 K to estimate the phonon contribution is shown in Fig.~\ref{fig:INSspecSM}. 
The results at low temperatures shown in Fig.~\ref{fig:INSspec} are also shown. 
Although the statistics at 250 K is worse than those of the data at low temperatures, we still observe the magnetic excitations corresponding to the 1--2 and 1--4 transitions at around 3 meV and 5 meV, respectively. 
The solid line on the 250 K data shows a fit with Eq.~(\ref{eq:SE}), where the energies and the width were fixed at $\Delta_{12}=3.1$ meV, $\Delta_{14}=5.2$ meV, and $\Gamma=1.2$ meV. 
If we fix the scale factor $C$ to the value at 10 K, the intensity of the magnetic scattering at 250 K is expected to be around the dashed line in the single-ion CEF model. 
Then, the phonon contribution to the intensity at 250 K is estimated to be around the dot-dashed line. 
If we correct for the Bose population factor, the phonon contribution at 50 K is estimated to be around the double-dot-dashed line. 
Even if we double this estimation, the phonon contribution at 50 K is expected to be much smaller than the observed intensity. 
Therefore, we can safely ascribe the observed intensity to the magnetic scattering.

\end{document}